\documentclass[reprint, amsmath, amssymb, aps]{revtex4-1}

\pdfoutput=1

\usepackage{graphicx}
\graphicspath{{./Figs/}}
\usepackage{dcolumn}
\usepackage{bm}

\usepackage{amsfonts}
\usepackage{amsmath,amssymb}
\usepackage{mathrsfs}

\usepackage{cancel}
\usepackage[dvipsnames]{xcolor}
\usepackage{enumerate}
\usepackage{lineno}
\usepackage[normalem]{ulem}

\usepackage[T1]{fontenc}
\usepackage{lmodern}

\definecolor{darkblue}{rgb}{0,0,0.9}
\usepackage{hyperref}
\hypersetup{
linktocpage,
colorlinks,
citecolor=darkblue,
filecolor=darkblue,
linkcolor=darkblue,
urlcolor=darkblue
}

\newcommand{\mc}{\mathcal}

\def\sla#1{\setbox0=\hbox{$#1$}\dimen0=\wd0
      \setbox1=\hbox{/} \dimen1=\wd1 \ifdim\dimen0>\dimen1
      \rlap{\hbox to \dimen0{\hfil/\hfil}} #1                        \else
      \rlap{\hbox to \dimen1{\hfil$#1$\hfil}}
      /   \fi}

\newcommand{\be}{\begin{equation}}
\newcommand{\ee}{\end{equation}}
\newcommand{\bea}{\begin{eqnarray}}
\newcommand{\eea}{\end{eqnarray}}

\newcommand{\nn}{\nonumber}

\def\parenbar#1{{\null\!      
   \mathop#1\limits^{\hbox{%
   \tiny{{\fontsize{3.5pt}{0em}\selectfont (}%
   \raisebox{-0.4pt}{--}%
   {\fontsize{3.5pt}{0em}\selectfont )}}}} 
   \!\null}} 

\DeclareOldFontCommand{\rm}{\normalfont\rmfamily}{\mathrm}
\DeclareOldFontCommand{\sf}{\normalfont\sffamily}{\mathsf}
\DeclareOldFontCommand{\tt}{\normalfont\ttfamily}{\mathtt}
\DeclareOldFontCommand{\bf}{\normalfont\bfseries}{\mathbf}
\DeclareOldFontCommand{\it}{\normalfont\itshape}{\mathit}
\DeclareOldFontCommand{\sl}{\normalfont\slshape}{\@nomath\sl}
\DeclareOldFontCommand{\sc}{\normalfont\scshape}{\@nomath\sc}

\begin{document}

\preprint{LAPTH-030/18}

\title{Effective-theory arguments \\ for pursuing lepton-flavor-violating $K$ decays at LHCb}


\author{Martino Borsato$^a$, Vladimir Vava Gligorov$^b$, Diego Guadagnoli$^c$,\\Diego Mart\'inez Santos$^a$, Olcyr Sumensari$^d$\\ \phantom{...}}

\affiliation{$^a$IGFAE, Universidade de Santiago de Compostela, Santiago de Compostela, Spain}

\author{}

\affiliation{$^b$LPNHE, Sorbonne Universit\'{e}, Universit\'{e} Paris Diderot, CNRS/IN2P3, Paris, France}

\author{}

\affiliation{
$^c$LAPTh, Universit\'{e} Savoie Mont-Blanc et CNRS, Annecy, France}

\author{}

\affiliation{$^d$Dipartimento di Fisica e Astronomia `G. Galilei', Universit\`a di Padova, Italy \\ Istituto Nazionale Fisica Nucleare, Sezione di Padova, I-35131 Padova, Italy
}

\begin{abstract}
\noindent We provide general effective-theory arguments relating present-day discrepancies in semi-leptonic $B$-meson decays to signals in kaon physics, in particular lepton-flavour violating ones of the kind $K \to (\pi) e^\pm \mu^\mp$. We show that $K$-decay branching ratios of around $10^{-12} - 10^{-13}$ are possible, for effective-theory cutoffs around $5-15$ TeV compatible with discrepancies in $B\to K^{(\ast)} \mu\mu$ decays. We perform a feasibility study of the reach for such decays at LHCb, taking $K^+ \to \pi^+ \mu^\pm e^\mp$ as a benchmark. In spite of the long lifetime of the $K^+$ compared to the detector size, the huge statistics anticipated as well as the overall detector performance translate into encouraging results. These include the possibility to reach the $10^{-12}$ ballpark, and thereby significantly improve current limits. Our results advocate LHC's high-luminosity Upgrade phase, and support analogous sensitivity studies at other facilities. Given the performance uncertainties inherent in the Upgrade phase, our conclusions are based on a range of assumptions we deem realistic on the particle identification performance as well as on the kinematic reconstruction thresholds for the signal candidates.
\end{abstract}

\maketitle


\noindent Data on $b \to s \ell \ell$ and $b \to c \ell \nu$ transitions display persistent deviations with respect to Standard-Model (SM) expectations \cite{Amhis:2016xyh,Altmannshofer:2017yso,Capdevila:2017bsm,Ciuchini:2017mik,DAmico:2017mtc,Hiller:2017bzc,Geng:2017svp}, suggesting a sizeable violation of Lepton Universality (LUV). Interestingly, the pattern of deviations finds a straightforward interpretation within an effective-field-theory (EFT) framework \cite{Altmannshofer:2017yso,Capdevila:2017bsm,Ciuchini:2017mik,DAmico:2017mtc,Hiller:2017bzc,Geng:2017svp}. Among the operator combinations able to explain at one stroke all the data, especially compelling from the ultra-violet standpoint is the product of two left-handed currents \cite{Hiller:2014yaa,Ghosh:2014awa,Bhattacharya:2014wla,Alonso:2014csa}. Since these interactions typically arise above the electroweak (EW) symmetry-breaking (EWSB) scale, fermions are in the `gauge' basis, that in general is misaligned with the mass-eigenstate basis. As a consequence, without further assumptions observable LUV is accompanied by Lepton-Flavor Violation (LFV), whose expected size is related to the measured amount of LUV \cite{Glashow:2014iga}. LFV may be expected in any $d \to d'$ transition, not only $b \to s$. In this work we present general arguments to relate LFV in $K$ decays to the existing LUV hints in $B$ decays, and produce predictions for the rates to expect. We then present a feasibility study on the reach for such $K$ decays at the upgraded LHCb experiment. This study aims at setting a realistic benchmark for the performance on such modes at the upgraded LHCb, given our present, limited knowledge of that phase of the experiment.

\medskip

\noindent {\bf Theory considerations --}  The most straightforward manifestation of LFV in kaon decays would be in $K \to (\pi) e \mu$ modes. Our aim is to relate predictions for these modes, that are mediated by the $s \to d$ current, to the present theory understanding of $B$-decay discrepancies, that occur in $b \to s$ transitions. In order to relate these two currents as model-independently as possible, we focus on an effective-theory picture, forgoing the introduction of new degrees of freedom.  To first approximation, such approach does not require any discussion of $b \to c$ discrepancies instead.

To illustrate our approach, let us first consider the 3rd-generation effective interaction
\be
\label{eq:HNP}
\mathscr{H}_{\rm NP} ~=~ G \, (\bar{b}'_L \gamma^\alpha b'_L)
(\bar{\tau}'_L \gamma_\alpha \tau'_L)~,
\ee
where $G \ll G_F$, $G_F$ is the Fermi constant, the subscript $L$ denotes left-handed fields, and primes identify the gauge basis \cite{Glashow:2014iga}. Below the EW scale this basis is related to the mass eigenbasis through chiral, unitary transformations, so that, even starting from the interaction (\ref{eq:HNP}), the new effects will in general propagate to generations other than the 3rd one.
Specifically, effects can be expected to also arise in decays of the kind $K \to (\pi) \ell \ell^{(\prime)}$, whether LFV or not. In fact, the first observable that comes to mind would be the $K$-physics analogue of $R_K$, namely the ratio $\mc B (K \to \pi \mu \mu) / \mc B (K \to \pi e e)$, which is, however, long-distance dominated \cite{D'Ambrosio:1998yj,Crivellin:2016vjc}. LFV Kaon decays instead, to which we focus our attention here, are null tests of the SM, and thereby free of any long-distance issue. The crucial question is whether they can be measurably large within SM extensions whose low-energy imprint is the interaction (\ref{eq:HNP}).
The first observation to be made is actually of experimental nature: limits on such modes are decade-old
\bea
\label{eq:limits}
\begin{tabular}{ccr}
$\mc B(K_L \to e^\pm \mu^\mp)$ &$< 4.7 \times 10^{-12}$ & \cite{Ambrose:1998us}~,\\
[0.1cm]
$\mc B(K_L \to \pi^0 e^\pm \mu^\mp)$ & $< 7.6 \times 10^{-11}$ & \cite{Abouzaid:2007aa}~,\\
[0.1cm]
$\mc B(K^+ \to \pi^+ e^- \mu^+)$ &$< 1.3 \times 10^{-11}$ & \cite{Sher:2005sp}~,\\ 
[0.1cm]
$\mc B(K^+ \to \pi^+ e^+ \mu^-)$ &$< 5.2 \times 10^{-10}$ & \cite{Appel:2000tc}~.
\end{tabular}
\eea
In order to translate Eq. (\ref{eq:HNP}) into general expectations for these modes, let us first rewrite it, {\em after EWSB,} as~\cite{Buttazzo:2017ixm}
\be
\label{eq:HNP_ZH}
\mathscr{H}_{\rm NP} ~=~ G \, \lambda_{ij}^q \lambda_{mn}^\ell \, (\bar d_i \gamma_L^\alpha d_j)(\bar \ell_m \gamma_{L \alpha} \ell_n)~,~~~G = \frac{C}{\Lambda^2}~,
\ee
where the UV scale $\Lambda$ is introduced for later convenience.
 The flavour structure of this theory is encoded in the (by construction) Hermitian $\lambda^{q,\ell}$ couplings, that will be discussed later on.
Besides, we will use the SM interaction $\mathscr H_{\rm eff}^{\rm SM} = N_{\rm SM} (\bar s_L \gamma^\mu u) (\bar \nu_L \gamma_\mu \mu) + {\rm H.c.}$, with $N_{\rm SM} =$ $4 G_F / \sqrt2 \cdot V_{us}^*$. Normalising the decay modes of interest so as to get rid of phase-space factors~\cite{Cahn:1980kv}, we find
\bea
\frac{\Gamma(K_L \to e^\pm \mu^\mp)}{\Gamma(K^+ \to \mu^+ \nu_\mu)} = \kappa^\ell \cdot \kappa^q_R ~~\left(= \frac{\Gamma(K_{S} \to \pi^0 \mu^\pm e^\mp)}{\Gamma(K^+ \to \pi^0 \mu^+ \nu_\mu)} \right)~,~~~~ \nn \\
[0.2cm]
\label{eq:LFV_K}
\frac{\Gamma(K_S \to e^\pm \mu^\mp)}{\Gamma(K^+ \to \mu^+ \nu_\mu)} = \kappa^\ell \cdot \kappa^q_I ~~\left(= \frac{\Gamma(K_{L} \to \pi^0 \mu^\pm e^\mp)}{\Gamma(K^+ \to \pi^0 \mu^+ \nu_\mu)} \right)~,~~~~ \\
[0.2cm]
\frac{\Gamma(K^+ \to \pi^+ \mu^\pm e^\mp)}{\Gamma(K^+ \to \pi^0 \mu^+ \nu_\mu)} = \kappa^\ell \cdot (\kappa^q_R + \kappa^q_I)~,~~~~\nn \hspace{2.45cm}
\eea
where we defined the abbreviations
\be
\label{eq:kappa}
\kappa^\ell \equiv \left| \frac{2G}{N_{\rm SM}} \right|^2 | \lambda^\ell_{12} |^2~, ~~ \kappa^q_R \equiv ({\rm Re}\lambda^q_{21})^2 ~, ~~ \kappa^q_I \equiv ({\rm Im}\lambda^q_{21})^2~.
\ee
These formulas hold under the excellent approximations of neglecting the electron mass and the mass differences between charged and neutral mesons, as well as $CP$ violation in mixing. We also note that the interaction in Eq.~(\ref{eq:HNP}) does not produce new tree-level contributions to the normalizing decays. The last members of Eqs.~(\ref{eq:LFV_K}), enclosed in parentheses, are quoted for completeness with respect to Eqs.~(\ref{eq:limits}). At LHCb, these modes pose a substantial additional challenge because of the final-state $\pi^0$ and will not be discussed further. 

\begin{figure}[!t]
\includegraphics[width = 0.495\textwidth]{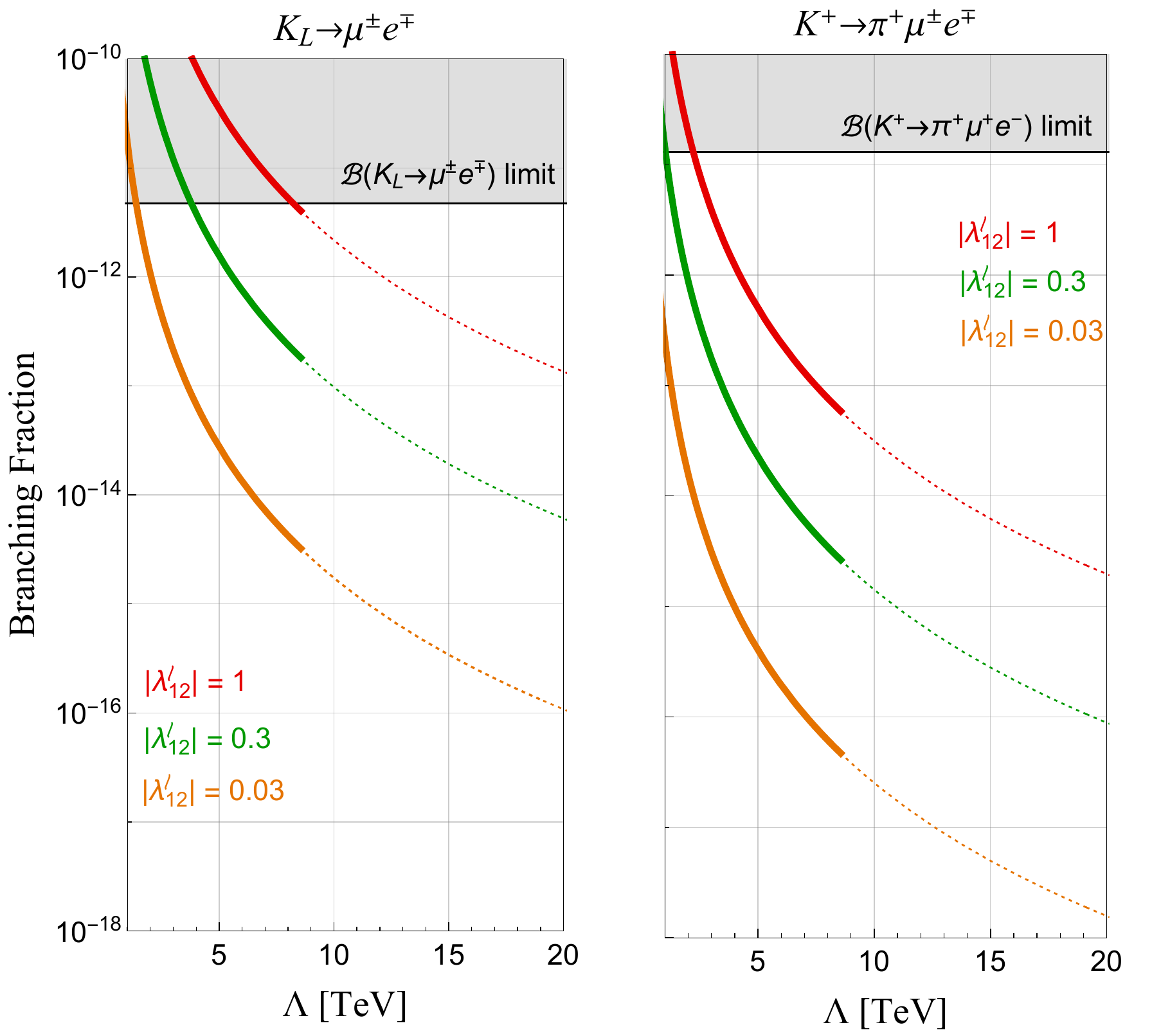}
 \caption{Predictions for $\mc B(K_L \to \mu^\pm e^\mp)$ and $\mc B(K^+ \to \pi^+ \mu^\pm e^\mp)$ as a function of the UV scale $\Lambda$.
Dashed lines signify that the parameter space is outside the $2\sigma$ range for $R_K$ \cite{Aaij:2014ora}. See text for further details.}
\label{fig:Kpred}
\end{figure}

Predictions for the modes in Eqs. (\ref{eq:LFV_K}) depend therefore on $\lambda^{q,\ell}_{12}$ and on the overall strength, $G$, of the new interaction. These three quantities can be constrained from the requirement that Eq.~(\ref{eq:HNP_ZH}) explain all relevant $B$-physics data, as we discuss next. 
First, departures from the limit $\lambda_{ij}^{q,\ell} = \delta_{i3} \delta_{j3}$ --~that yields back Eq.~(\ref{eq:HNP})~-- may be parameterised by the spurions of a suitably chosen, global flavour symmetry \cite{Barbieri:2015yvd}; $B$-physics anomalies can then be accounted for by appropriate ranges for $C/\Lambda^2$ and for the spurions parameterizing the relevant $\lambda^{q,\ell}$ entries, as discussed for example in Ref. \cite{Feruglio:2017rjo}. Concerning $\lambda^q$, an efficient approach is a CKM-like ansatz
\be
\label{eq:lambda_q}
\lambda^q_{ij} = b_q V_{ti}^* V_{tj}~,
\ee
with $V$ the Cabibbo-Kobayashi-Maskawa (CKM) matrix and $b_q$ a flavour-blind coupling
\cite{Buttazzo:2017ixm}.
We will adhere to this ansatz, suggested in particular by the constraints imposed by data on Atomic Parity Violation \cite{Langacker:1990jf,Dzuba:2012kx} as well as $\mu \to e$ conversion in nuclei \cite{Dohmen:1993mp,Bertl:2006up}. Our coupling of interest is then fixed as $\lambda^q_{21} = b_q V_{ts}^* V_{td}$ \footnote{For definiteness we took $|V_{ts}| = 4.0 \times 10^{-2}$, $|V_{td}| = 8.6 \times 10^{-3}$ from a `tree-level' CKM fit \cite{Derkach-private,UTfit,CKMfitter}.}, that amounts to a suppression mechanism for the effects we are seeking to predict. For the lepton-sector couplings $\lambda^\ell$ there is larger freedom,
because of the model-building uncertainties inherent in the lepton sector. We will accordingly adopt an agnostic approach, and discuss predictions with hierarchically different values for $|\lambda^\ell_{12}|$  (see legend of Fig. \ref{fig:Kpred}). In spite of this freedom, we will see that bounds on $b \to s \mu e$ modes are constraining enough that our approach stays predictive.

Besides the flavourful, channel-specific couplings just discussed, our relevant amplitudes depend on the choice of the product $\bar C \equiv C \cdot b_q$  of two flavour-blind numbers, namely the overall strength $C$ of the interaction Eq.~(\ref{eq:HNP_ZH}) as well as the normalization $b_{q}$ of the $\lambda^{q}$ coupling matrix Eq.~(\ref{eq:lambda_q}).
Given the normalizations in Eqs.~(\ref{eq:HNP_ZH}) and (\ref{eq:lambda_q}), 
the $\bar C$ coupling will be at most around unity or $4\pi$ for a perturbative or respectively non-perturbative UV theory. We will display predictions for one reference value: $\bar C = 1$, and add comments where appropriate. We note that predictions assuming, say, $\bar C = 4\pi$, can be obtained by trivially multiplying by $(4 \pi)^2$ those at $\bar C = 1$.

Eqs.~(\ref{eq:LFV_K}) translate into the following predictions for our modes of interest \footnote{We use $\mc B(K^+ \to \mu^+ \nu_\mu) \simeq 63.6\%$, $\mc B(K^+ \to \pi^0 \mu^+ \nu_\mu) \simeq 3.35\%$, $\Gamma(K^+) / \Gamma(K_L) \simeq 4.13$ and $\Gamma(K_L) / \Gamma(K_S) \simeq 1.75 \times 10^{-3}$ \cite{Olive:2016xmw}.}
\bea
\label{eq:BRKpheno}
&&\mc B(K_{L} \to \mu^\pm e^\mp) \simeq 2.6 \, \kappa^\ell \kappa^q_R~,\nn \\
&&\mc B(K_{S} \to \mu^\pm e^\mp) \simeq 4.6 \cdot 10^{-3} \, \kappa^\ell \kappa^q_I~, \\
&&\mc B(K^+ \to \pi^+ \mu^\pm e^\mp) \simeq 0.034 \, \kappa^\ell (\kappa^q_R +\kappa^q_I)~.\nn
\eea
Assuming the $\kappa^q_I$ coupling to be comparable to $\kappa^q_R$, the LFV mode of the $K_S$ is suppressed by a factor of about $\Gamma_{K_L} / \Gamma_{K_S} \simeq 1.75 \times 10^{-3}$ with respect to the corresponding $K_L$ mode. Notably, this physics suppression factor is, at LHCb, nearly compensated by the experimental acceptance enhancement, so that the product is invariant. We will comment further on these two modes at the end of the next Section.

In Fig.~\ref{fig:Kpred} we display predictions for $\mc B(K_L \to \mu^\pm e^\mp)$ (left panel) and $\mc B(K^+ \to \pi^+ \mu^\pm e^\mp)$ (right panel) versus the new-physics scale $\Lambda$ in the normalization of Eq.~(\ref{eq:HNP_ZH}). The color code refers to three possible choices for the leptonic coupling $\lambda^\ell_{12}=(\lambda^\ell_{21})^\ast$. Solid vs.~dashed lines represent predictions in agreement with, and respectively outside, the 2$\sigma$ range for $R_K^{(\ast)}$ \cite{Aaij:2014ora}. These constraints impose the upper bound $\Lambda \lesssim 8.6$ TeV (end of solid lines), under the assumption $\bar C = 1$, representative of a perturbatively coupled UV theory.
(A larger $\bar C$, as in strongly-coupled new-physics scenarios, would increase the corresponding upper bound on $\Lambda$ accordingly.) We note that the $R_K^{(\ast)}$ constraints depend on $\lambda^\ell_{22}$, and the mentioned bound on $\Lambda$ arises from the requirement $|\lambda^\ell_{22}|<1$.
In short, the $R_K^{(\ast)}$ constraints tend to push towards the left of Fig.~ \ref{fig:Kpred}, as one may intuitively expect.

A crucial constraint for our $K \to (\pi) \mu e$ predictions in Fig.~\ref{fig:Kpred} are the existing limits on $b \to s \mu e$ modes, in particular $\mc B(B \to K \mu^\pm e^\mp) < 3.8 \times 10^{-8}$~\cite{Aubert:2006vb}, $\mc B(B_s^0 \to \mu^\pm e^\mp) < 1.1 \times 10^{-8}$~\cite{Aaij:2013cby}, $\mc B(B \to K^* \mu^\pm e^\mp) < 1.8 \times 10^{-7}$~\cite{Sandilya:2018pop}. Relevant formulas are implemented following Ref.~\cite{Becirevic:2016zri}.
These limits imply $|\lambda^\ell_{12}| \lesssim 0.028$, imposed particularly by the first of the above modes. This limit, represented by the orange lines in Fig.~\ref{fig:Kpred}, in turn translates into the upper bounds $\mathcal{B}(K_L \to \mu^\pm e^\mp)\lesssim 1.7 \times 10^{-11}$ and $\mathcal{B}(K^+ \to \pi^+ \mu^\pm e^\mp)\lesssim 2.5 \times 10^{-13}$. As discussed, such bounds are obtained with a CKM-like ansatz for the relevant quark coupling $\lambda^q_{21}$ -- see Eqs. (\ref{eq:kappa})-(\ref{eq:lambda_q}) -- that holds up to a factor of $O(1)$. Even taking into account this freedom, we can safely conclude that the scenario represented in Fig.~\ref{fig:Kpred} by the red lines is excluded by the mentioned $b \to s \mu e$ modes.

In summary, we obtain predictions for LFV $K$ branching ratios that may 
realistically be around $5 \times 10^{-13}$ for $K^+$ modes, and one 
order of magnitude above for the $K_L$. 
Such figures are quite encouraging, taking into account the discussed, severe parametric suppressions imposed by existing constraints
\footnote{One may start from a completely different stance, and e.g. consider flavour models for the chiral transformations relating the primed basis in Eq. (\ref{eq:HNP}) to the mass eigenbasis.
Examples include \cite{Guadagnoli:2015nra}, motivated by a solution to the strong-$CP$ problem, or \cite{Boucenna:2015raa}, which is an attempt to connect $B$-physics anomalies to neutrino physics. In both cases one obtains, again, LFV $K$-decay branching ratios in the ballpark of $10^{-13}$.}.

\medskip

\noindent {\bf LHCb reach --} We next discuss the LHCb reach for the above mentioned kaon decays as a function of the integrated luminosity to be collected by the LHCb experiment and its upgrades~\cite{Bediaga:2012uyd, Aaij:2244311}. We parametrize the differential cross section for kaon production in 13-TeV $pp$ collisions using Pythia~8.230~\cite{Sjostrand:2006za,Sjostrand:2007gs} with default tuned parameter set. We obtain average $K^+$ rapidity densities in agreement within 5\% with CMS measurements at 0.9, 2.76, 7 and 13~TeV~\cite{Chatrchyan:2012qb, Sirunyan:2017zmn} for central rapidity, thus validating the Pythia description. At 13~TeV we obtain a total $K^\pm$ cross section of 0.63 barn that, in the LHCb pseudorapidity acceptance $2<\eta<5$, translates into a $K^\pm$ cross section as large as $0.14$~barn. From such encouraging figure, and using the discussed theoretical expectations, we then proceed to study the reach for kaon LFV modes at the LHCb upgrades, taking $K^+\to\pi^+ \mu^\pm e^\mp$ as a benchmark.

We estimate the LHCb detector response using the RapidSim package~\cite{Cowan:2016tnm}, which implements a parametric simulation of the LHCb detector acceptance, momentum, and vertex resolutions, including electron bremsstrahlung. The assumed performances, that we describe in detail next, are in line with the `standard' assumptions made about Upgrade II in LHCb literature, including the recent Upgrade II physics case \cite{Bediaga:2018lhg}.

The default RapidSim parametrization of momentum and vertex resolutions -- both critical for the kaon-mass resolution -- are tuned specifically for this analysis using public LHCb numbers for kaon decays~\cite{Aaij:2017ddf}. Besides, to get accurate estimates of the acceptance we perform an approximate simulation of the LHCb-upgrade tracking system using \cite{Collaboration:1647400,Collaboration:1624070} (see also~\cite{Hommels:999327} for the magnetic-field modelling).

The $K^+\to \pi^+\mu^\pm e^\mp$ candidate is considered if all of its decay products lie within the LHCb tracker acceptance, leave hits in both the vertex detector and tracker stations, and if each daughter particle crosses at least three stations in the vertex detector. This last criterion imposes a roughly 0.5m-long decay volume. By comparing the efficiencies we obtain for $K_{\rm S}\to\pi^+\pi^- e^+ e^-$ to LHCb public numbers in~\cite{MarinBenito:2193358}, we estimate that electrons from kaon decays are $50\%$ less likely than pions or muons to be reconstructed and we correct our simulation accordingly.

Because of the light kaon mass, the selection criteria dominating the signal efficiency are the kinematic requirements on the final-state decay products, most notably the kinematic threshold to reconstruct charged-particle tracks in LHCb's real-time processing (trigger). LHCb's upgrade trigger will have access to all information from all of LHCb's subdetectors at the full LHC collision rate. Therefore, the aforementioned kinematic threshold will realistically be limited by computing resources rather than inherent detector limitations. While it is hard to make dependable predictions with present knowledge, to make progress we next discuss a few assumptions we deem reasonable. 
First, we demand a momentum larger than 2~GeV for all charged tracks so that they are not swept out of the detector acceptance by the dipole magnet; muon candidates are required to have a momentum in excess of 3~GeV in order to reach the muon stations.
Then, we foresee a trigger strategy involving the identification of a muon track in the muon stations, which is subsequently matched to the vertex detector and upstream tracking stations, and required to have a large impact parameter with respect to the collision vertex. We expect such strategy will allow to reconstruct muons in real time down to 0.1~GeV in transverse momentum.
The muon track could then be used to identify a region of interest in the tracker where two further displaced tracks are looked for to form a $K^+\to \pi^+\mu^\pm e^\mp$ candidate. We assume that the $\pi^+$ and $e^\mp$ candidate tracks can be reconstructed in real time if their transverse momentum, $p_{\rm T}$, exceeds a threshold value that we vary between 0.1 and 0.3~GeV.
 
We next discuss signal separation. Thanks to the long lifetime of charged kaons, selection criteria on the impact parameters of final-state tracks allow, as a rule, to greatly reduce the combinatorial background from tracks coming from the $pp$ collision vertex. However, at the luminosity expected at LHCb Upgrade II, $O(10^{34})~{\rm cm}^{-2}{\rm s}^{-1}$, it is difficult to reliably estimate this background, which could get a sizable contribution also from fake tracks due to random associations of hits in the trackers caused by the huge luminosity itself. In this study we assume these backgrounds will be negligible with respect to those coming from misidentified (mis-ID) kaon decays, which we estimate by simulating the processes reported in Table~\ref{tab:simproc} with RapidSim.
\begin{table}[]
\centering
\begin{tabular}{c|c|c}
Decay & BR & mis-ID \\\hline
$K^+\to \pi^+\pi^+\pi^-$ & $5.6\times 10^{-2}$ & $\pi^+ \to \mu^+$ and $\pi^- \to e^-$ \\
$K^+\to \pi^+\mu^+\mu^-$ & $0.94\times 10^{-7}$ & $\mu^- \to e^-$\\
$K^+\to \pi^+ e^+e^-$ & $3.0\times 10^{-7}$ & $e^+ \to \mu^+$
\end{tabular}
\caption{Backgrounds from final-state mis-identification. Branching ratios are taken from~\cite{Patrignani:2016xqp}.}
\label{tab:simproc}
\end{table}
We estimate their abundances in the $m(\pi^+\mu^\pm e^\mp)$ signal region based on their known branching fractions, their mass spectra obtained from our tracking parametrization, and PID performances estimated from~\cite{Collaboration:1624074, Aaij:2014azz}. We obtain total mis-ID background yields at 300~fb$^{-1}$ (for tracks $p_{\rm T}>0.1$~GeV) between 5 and 900 events, and we label these two values as optimistic vs. pessimistic scenarios for LHCb-Upgrade-II PID performances.
More details about our procedure can be found in the Appendix.

\begin{figure}[!t]
\centering \includegraphics[width = 0.495\textwidth]{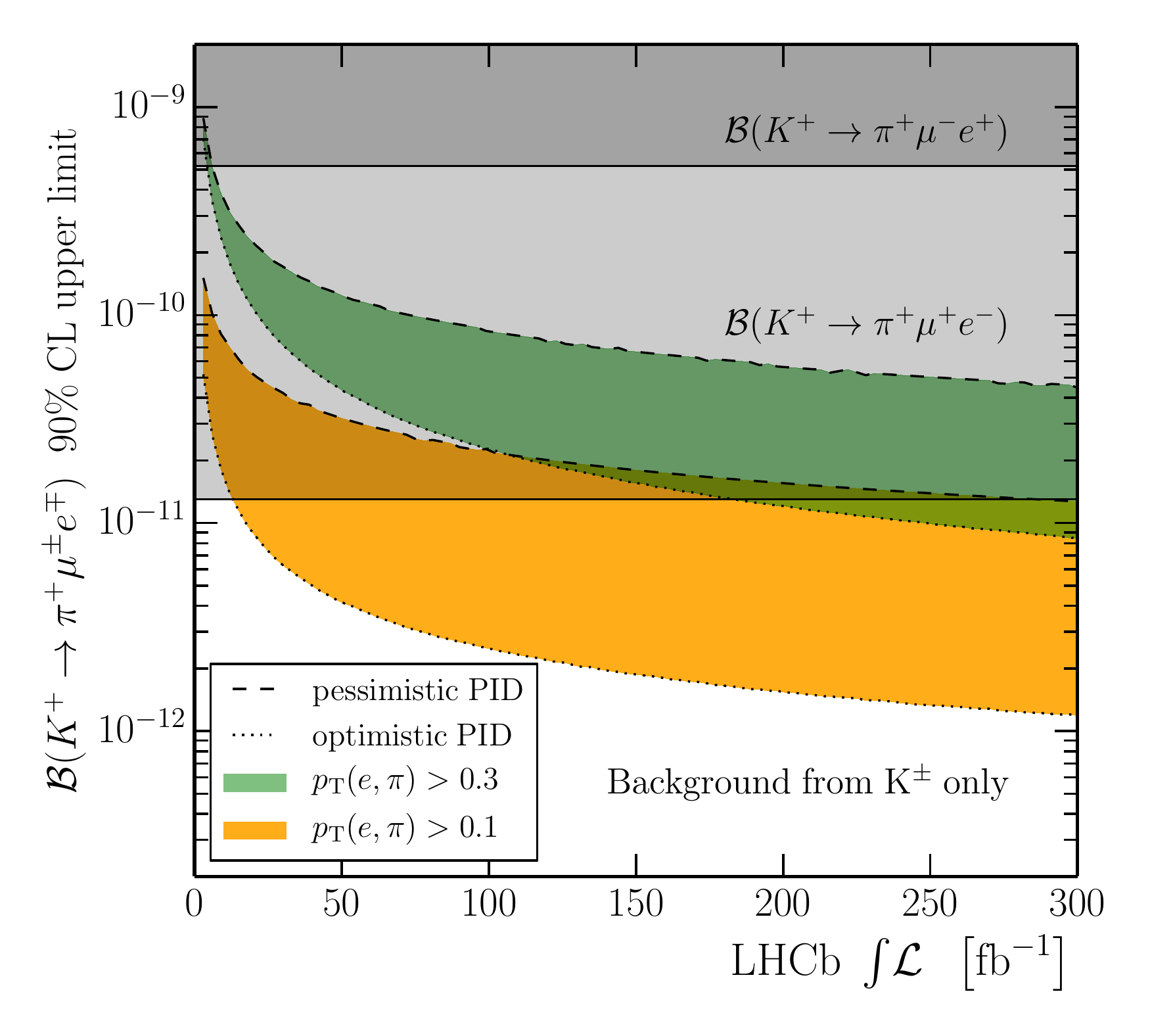} \hfill
\vspace{-0.5cm}
 \caption{LHCb expected reach in terms of expected 90\% upper limit on $K^+\to \pi^+\mu^\pm e^\mp$ as a function of the integrated luminosity with 13~TeV $pp$ collisions. Different scenarios in terms of PID performance and $p_{\rm T}$ thresholds of the $\pi^+$ and $e^{\pm}$ candidates are shown. Possible backgrounds from combinatorial and ghost tracks have not been considered.}
\label{fig:BRvslumi}
\vspace{-0.3cm}
\end{figure}
The estimated background mass spectra are used to obtain background yields in a signal mass region between 0.480 and 0.505~GeV. From them, a counting-experiment approach is used to obtain the expected $90\%$ confidence level upper limits on the $K^+\to \pi^+\mu^\pm e^\mp$ branching ratio. The upper limits are shown as a function of the integrated luminosity and for different scenarios of detector performance in Fig.~\ref{fig:BRvslumi}.
The figure shows that LHCb has the potential to probe branching fractions between $10^{-12}$ and $5\times 10^{-11}$ with $300~{\rm fb}^{-1}$ of integrated luminosity.

We emphasize that LHCb will also be able to probe $\parenbar{K}{}^0 \to \mu^\pm e^\mp$, whether the initial state belongs to a $K_S$ or to a $K_L$. The LHCb acceptance is roughly 100 times better for $K_S^0$ than for $K^\pm$, and roughly 3 times worse for $K_L^0$ than for $K^\pm$. This implies a $K^0_S / K^0_L$ acceptance improvement compensating almost exactly the relative lifetime suppression, as already commented on below eq.~(\ref{eq:BRKpheno}). This compensation is not accidental. In fact, in the limit of a very small VELO, it can be shown that the acceptance ratio is indeed $\tau_L/\tau_S$. It is beyond any doubt that LHCb can produce a world-best measurement of $\mathcal{B}(K_S^0\to\mu^\pm e^\mp)$ since it would be the first search for this decay. But, interestingly, assuming a similar performance and background level as we did for $K^+\to \pi^+\mu^\pm e^\mp$, even a competitive measurement of $\mathcal{B}(K_L^0\to\mu^\pm e^\mp)$ might be feasible. Keeping in mind that the $K^0_L$ and $K^0_S$ decay modes are sensitive to the real and respectively the imaginary part of $\lambda^q$, see Eqs. (\ref{eq:LFV_K}), a joint measurement of the two modes would serve as a model discriminator.

In conclusion, we presented general, effective-theory arguments that relate existing signals of LUV in $B$ decays to possible signatures in $K$ physics, focusing on LFV decays, that are free from long-distance SM contributions.
These arguments rest on the main assumption of a $(V-A)\times(V-A)$ 4-fermion interaction coupled mainly to the 3rd generation (in the gauge basis) and on a CKM-like structure for the flavourful quark couplings, whereas we stay agnostic on the lepton couplings.
We obtain predictions for $\mc B(K_L \to \mu^\pm e^\mp)$ right beneath the existing limit of $4.7 \times 10^{-12}$, and for $\mc B(K^+ \to \pi^+ \mu^\pm e^\mp)$ in the range $10^{-12} - 10^{-13}$, if the new-physics scale is relatively light, $\lesssim 10$ TeV. We performed a sensitivity study of these modes at the LHCb in its upgrade phase, taking the $K^+ \to \pi^+ \mu^\pm e^\mp$ mode as a benchmark. With a range of motivated assumptions \cite{Bediaga:2018lhg} for all the known unknowns (including the kinematic thresholds to reconstruct charged particles in real time and the PID performance), we find that LHCb may update {\em all} the existing limits, and probe a sizable part of the parameter space suggested by the $B$-physics discrepancies.

The main message of our study is that LHCb, an experiment not explicitly designed for kaon decays, may well be very competitive in the context of rare and LFV kaon decays. This conclusion in turn calls attention to other running and upcoming facilities including NA62~\cite{NA62:2312430} and the newly proposed TauFV~\cite{PBCTalk} experiment. NA62 is a dedicated $K^+$ experiment with exquisite light-lepton identification capabilities. According to crude estimations, it could reach the $10^{-12}$ ballpark in LFV kaon decays~\cite{Petrov:2017wza} with the data collected so far.
The TauFV experiment may benefit from no less than $O(10^{19})$ kaons in a decay volume of a similar size to LHCb's and with a similar detector layout. We hope that our results will encourage dedicated sensitivity studies for these facilities.

\begin{acknowledgments}
\noindent The authors would like to thank Johannes Albrecht, Mat Charles, Francesco Dettori, Tim Gershon, and Guy Wilkinson for useful discussions and comments on a draft of this paper. DG acknowledges useful exchanges with Andrzej Buras and Dario Buttazzo. The work of DG is partially supported by the CNRS grant PICS07229. The work of MB and DMS is supported by ERC-StG-639068  ``BSMFLEET''. The work of OS is supported by the European Union's Horizon 2020 research and innovation programme under the Marie Sklodowska-Curie grant agreement N$^\circ$~674896. The work of VVG is partially supported by ERC-CoG-724777 ``RECEPT''.
\end{acknowledgments}

\section*{Appendix}

\begin{figure*}[t]
\centering 
\includegraphics[width = 0.49\textwidth]{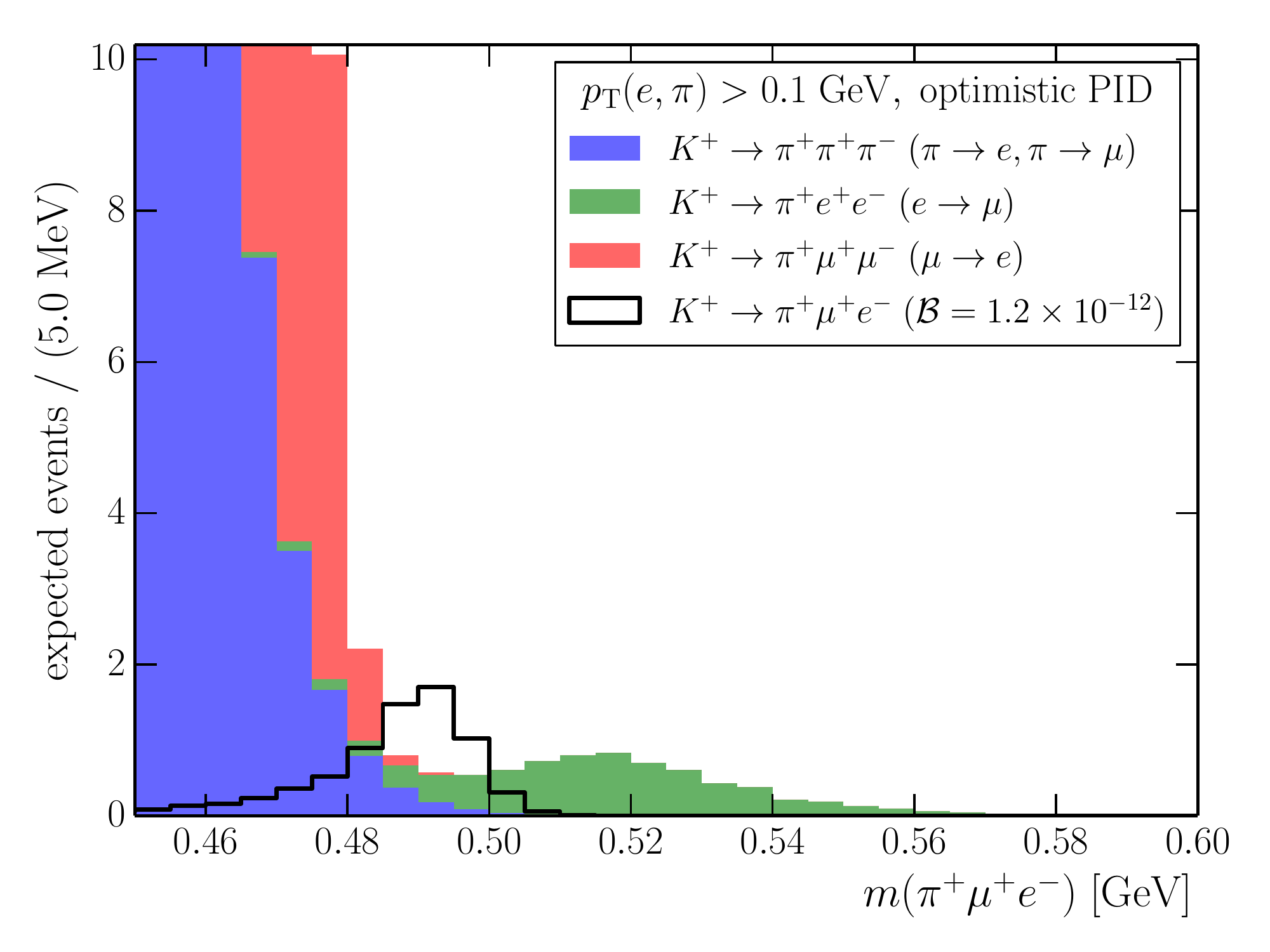} 
\includegraphics[width = 0.49\textwidth]{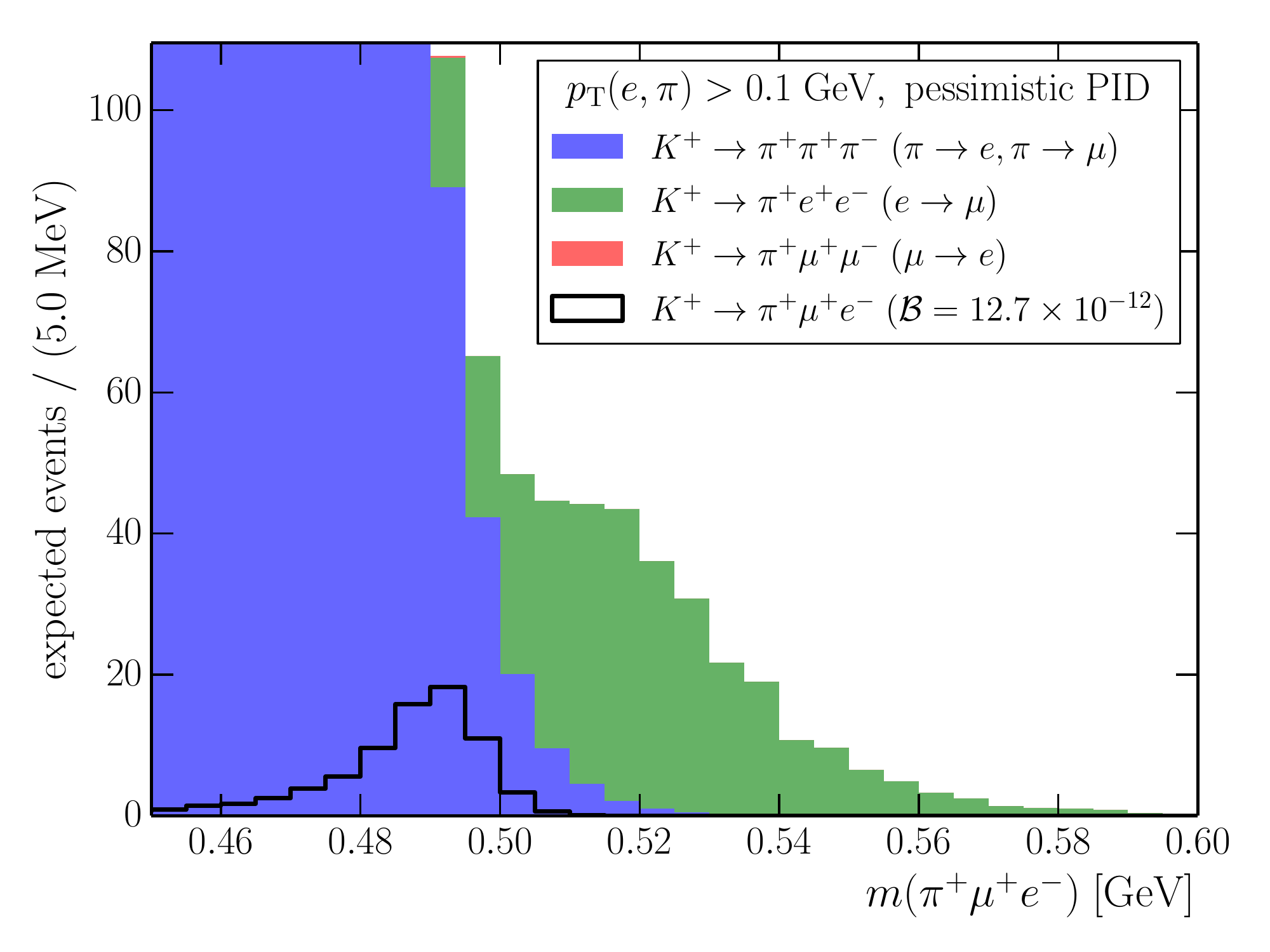}
\includegraphics[width = 0.49\textwidth]{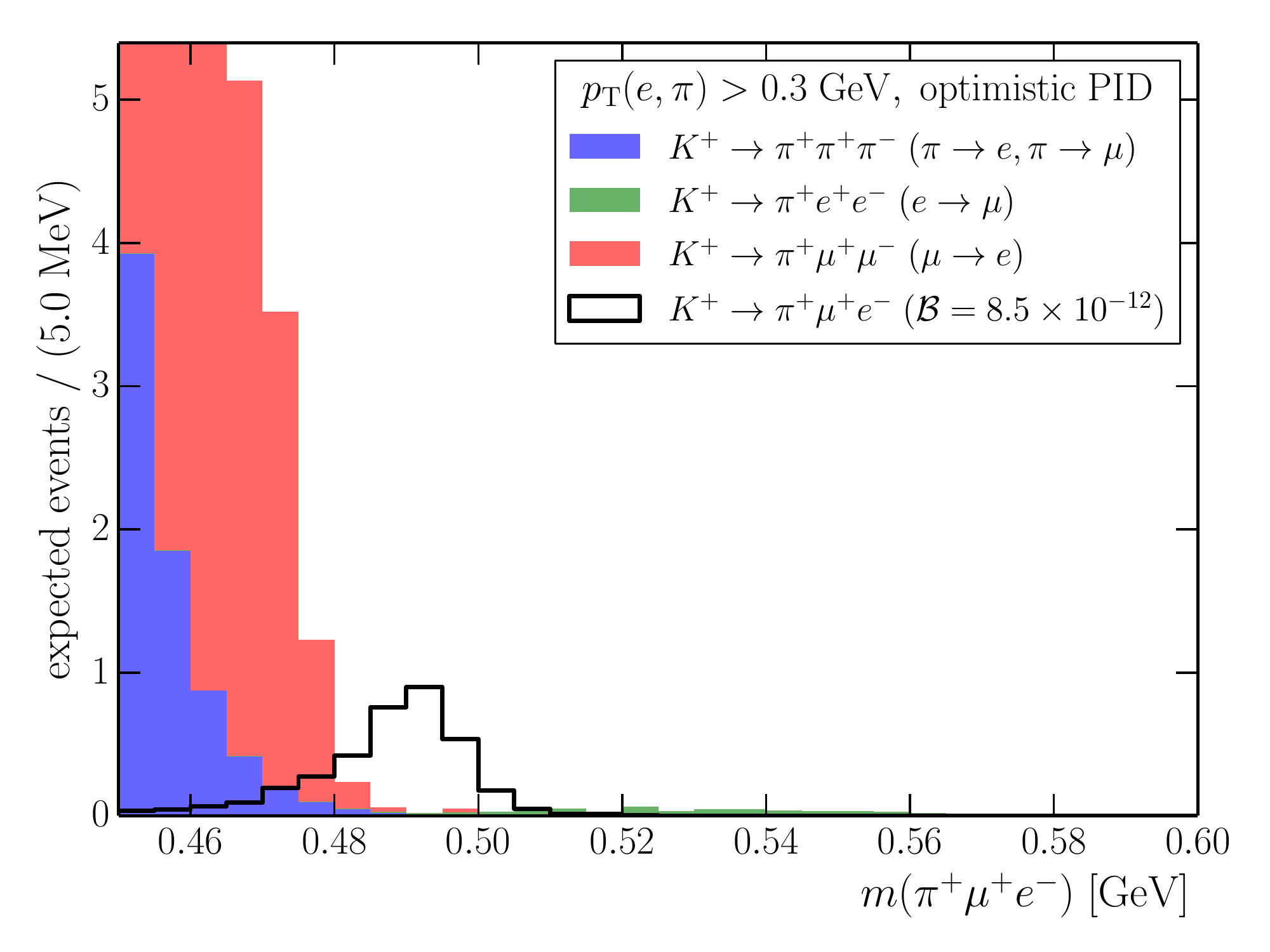} 
\includegraphics[width = 0.49\textwidth]{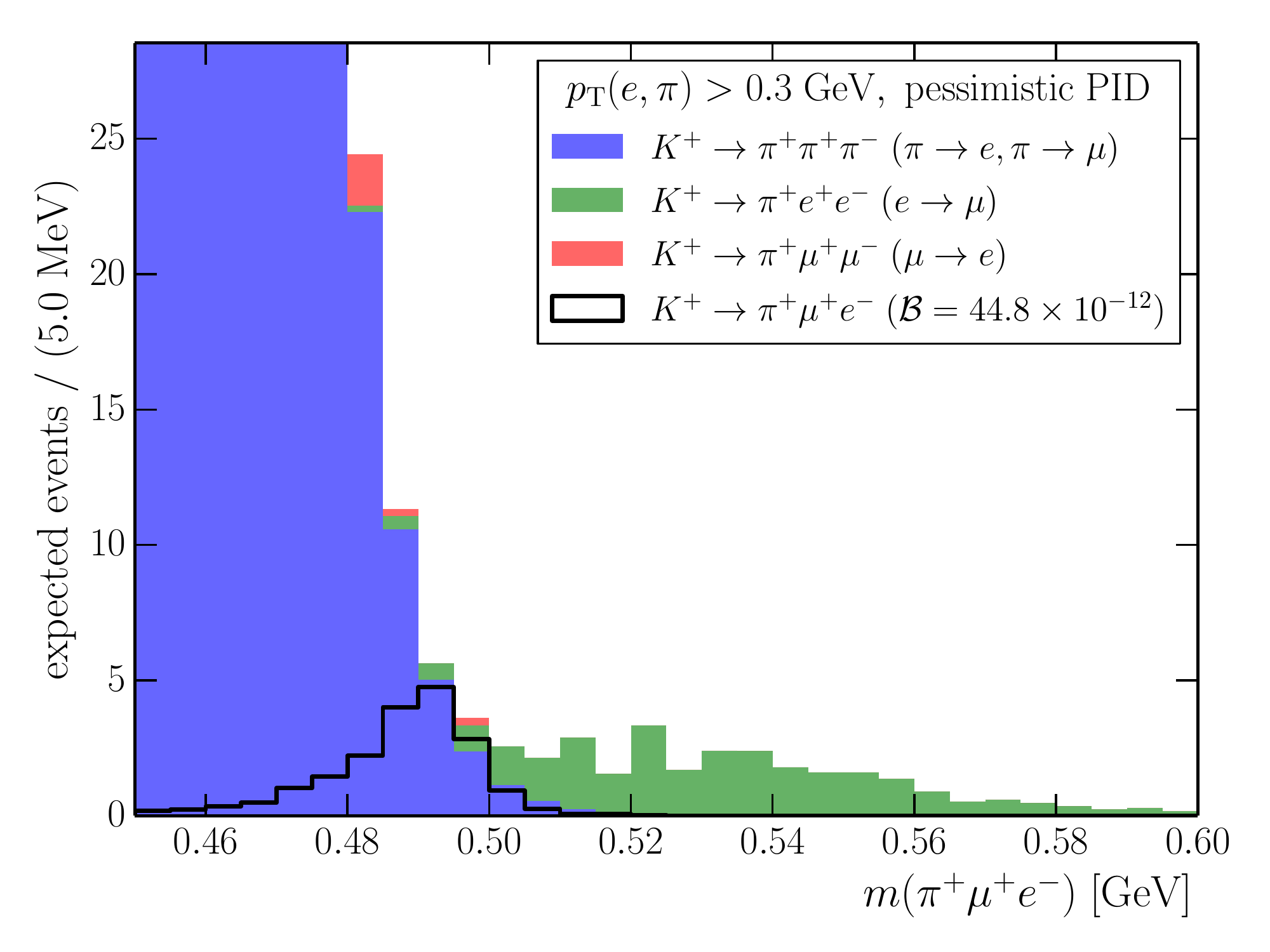}
 \caption{Mass spectra expected at LHCb for an integrated luminosity of 300~fb${ }^{-1}$ of 13~TeV $pp$ collisions. Four different scenarios in terms of PID performance and $p_{\rm T}$ threshold are shown. The signal branching ratio shown for each scenario is the one of the corresponding expected limit.
}
\label{fig:massplots}
\end{figure*}

\noindent Since the particle-identification performance of LHCb's upgrades may be significantly different than the current detector's, we study a range of misidentification working points, as reported in Table~\ref{tab:misid}. The mass spectra corresponding to our different performance assumptions are shown in Fig.~\ref{fig:massplots}.
\begin{table}[b]
\centering
\begin{tabular}{c|c|c}
mis-ID & optimistic  & pessimistic    \\
\hline
$\pi^\pm$ mis-ID as $\mu^\pm$ & $0.1\%$ & $5\%$ \\
$\pi^\pm$ mis-ID as $e^\pm$   & $0.1\%$ & $1\%$ \\
$e^\pm$ mis-ID as $\pi^\pm$   & $0.1\%$ & $1\%$ \\
$e^\pm$ mis-ID as $\mu^\pm$   & $0.01\%$ & $0.05\%$ \\
$\mu^\pm$ mis-ID as $e^\pm$  & $0.01\%$ & $0.1\%$ \\
\end{tabular}
\caption{Ranges of misidentification probabilities assumed in this study for efficiencies of correct identification of $50\%$ for electrons and $90\%$ for muons and pions.}
\label{tab:misid}
\end{table}
\\
Electrons are reconstructed with a median momentum between 5 and 10~GeV, depending on the trigger $p_{\rm T}$ threshold. In this momentum range, the RICH detectors provide high discriminating power with respect to pions and muons, as the majority of $\pi$ and $\mu$ tracks will have much smaller Cherenkov rings than electrons or will not emit Cherenkov light at all.
Pions in this kinematic range can be misidentified as electrons with up to $1\%$ probability~\cite{Collaboration:1624074} for an electron identification efficiency of 50\%. \\
Owing mainly to the muon chambers, pions can be misidentified as muons with up to $5\%$ probability~\cite{Collaboration:1624074} (assuming $90\%$ identification efficiency) in the relevant range of momentum around 6~GeV.\\
Electrons misidentified as pions are rare (and this is even more so for muons) thanks to the different signatures they have in the RICH detectors, as already mentioned, as well as in the calorimeters and muon chambers. We estimate the respective PID performance by multiplying the probabilities of $\mu^\pm$ mis-ID as $\pi^\pm$ and of $\pi^\pm$ mis-ID as $e^\pm$, which yields $5\times 10^{-4}$. This probability may be reduced with a dedicated optimisation, but such analysis would not help for branching ratios below $10^{-5}$.\\
Finally, we expect a small fraction of muons mis-ID as electrons, which we vary between $10^{-4}$ and $10^{-3}$.\\
Significantly better performances than in our pessimistic scenario may well be possible thanks to the combination of all available information through machine-learning techniques (\emph{e.g.} $10^{-3}$ rejection is reached in~\cite{Aaij:2014azz}), as well as with future optimisation and the kind of higher-granularity electromagnetic calorimeter being studied for future upgrades of LHCb~\cite{Aaij:2244311}. Any of such improvements, although likely, is however difficult to quantify at present.

\bibliography{note-Kdecays-LHCb}
\clearpage

\end{document}